\documentclass[aps,twocolumn,pra,showpacs,floatfix]{revtex4}
\usepackage{epsfig}
\usepackage{graphicx}
\usepackage{dcolumn}
\usepackage{amssymb,amsmath}
\usepackage{mathrsfs}
\begin{document}

%\documentclass[aps,pra,twocolumn,nofootinbib]{revtex4-2}
%\usepackage[colorlinks]{hyperref}
%\usepackage[utf8]{inputenc}
%\usepackage{graphicx}
%\usepackage{multirow}
%\usepackage{amssymb}
%\usepackage{amsmath}
%\usepackage{microtype}
%\begin{document}
	
\title{Nuclear polarisation and relativistic effects contributions to King plot non-linearity}%%% ????? %%%
	
%\author{}
%\affiliation{}
\author{P. Munro-Laylim, V. A. Dzuba, and V. V. Flambaum}

\affiliation{School of Physics, University of New South Wales, Sydney 2052, Australia}

\begin{abstract}
The effect of nuclear polarisation on the non-linearities of King plot for isotope shift in Ca$^+$,  Sr$^+$ and  Yb$^+$ ions is studied.
These ions are currently used in search for the manifestation of new interaction via non-linearities of King plot.
%It is demonstrated that the nuclear polarisation effect is significant and should be taken into account in the analysis. %In heavy atoms or ions with deformed nuclei the effect of deformation dominates over nuclear polarisation.
 The role of relativistic effects is also studied. %It is demonstrated that their contributions to the non-linearities in spherical nuclei  is very small. 
 The best way to achieve  separation of electronic and nuclear variables (which is necessary for the linearity) is to present the field shift as $F''\delta \langle r^{\lambda} \rangle$, where the value of $\lambda$  is the result of numerical relativistic calculations ($2\gamma < \lambda <2$, $\gamma = \sqrt{1-(Z/137.036)^2}$).  Using this ansatz we demonstrated  that the contribution of the relativistic effects to the non-linearity in spherical nuclei is very small.
\end{abstract}
	
\maketitle
	
\section{Introduction}\label{SectionIntroduction}
The Standard Model (SM) has been one of the most successful theories in physics, proving to be accurate in its numerous predictions. However, significant flaws still arise in the SM, including the lack of an explanation for dark matter and matter-antimatter asymmetry, among other problems. The apparent incompleteness leads us to explore for physics beyond the SM, with a variety of avenues being tested to find new physics. One particular area of exploration is with isotope shift (IS), where significant improvements in measurements have been made in recent years  - see e.g. \cite{CountsPRL2020,KnollmannPRA2019,MiyakePRR2019,ManovitzPRL2019,Yb+Yb}.  Within SM, the  King plot for isotope shift  \cite{King1963} is approximately linear, up to some small contributions.  However, in Refs. \cite{CountsPRL2020,MiyakePRR2019,Yb+Yb} small deviations  from linearity in the King plot have been  observed. Multiple explanations have been offered in Refs. \cite{DelaunayPRD2017,BerengutPRL2018,FlambaumPRA2018} to account for the observed non-linearity, including higher order correction terms to field isotope shift (FIS) within SM, but a possibility of a new interaction, mediated by a boson beyond  SM, is also a valid explanation for the observations.

It was proposed in in Refs. \cite{DelaunayPRD2017,BerengutPRL2018,FlambaumPRA2018}  that this boson $\phi$ acts as a mediator between neutrons and electrons, which may contribute a Yukawa potential to the nuclear potential. Due to this interaction  the boson affects FIS, however such a small contribution would be impossible to detect among SM contributions in direct IS measurements. This Yukawa potential  would be more apparent in the King plots due to its contribution to the non-linearity. %This boson may also couple to protons, however such an interaction does  not contribute to the King plot non-linearity due to lack of isotopic dependence. 
This effect on the King plot non-linearity has been examined in Ref. \cite{CountsPRL2020,FrugiuelePRD2017,RehbehnPRA2021}, however these explorations neglected SM contributions to the  non-linearities; while the work is notable, findings cannot be conclusive as  true constraints on the boson. Therefore, the importance of accurately determining SM contributions to the King plot non-linearity is apparent, as this will allow future works to apply more accurate bounds on the new boson.
 %%% Discuss this further?
% Require reference for 'cannot detect in regular IS, need King plots', cannot recall which paper

The non-linearities of King plot arise from small SM  contributions to  FIS, which include  quadratic field shift, relativistic effects and effects of nuclear deformation.
%can be presented in the form for a atomic transition $a$
%\begin{equation}
%    \nu_a = F_a \delta\langle r^2 \rangle + G_a^{(2)}\delta\langle r^2 \rangle^2 + G_a^{(4)}\delta\langle r^4 \rangle,
%\end{equation}
%wherein the first term is the first order of FIS contributing to linearity, and the second and third terms represent the quadratic field shift (QFS), i.e. second order corrections to the change in nuclear potential, and relativistic and nuclear deformation effects, respectively. The latter two effects contribute to King plot non-linearity and have been shown to have a similar contribution in Ref. \cite{AllehabiPRA2021}. Also in this paper, it is presented how the $G^{(2)}$ and $G^{(4)}$ coefficients can be calculated, allowing non-linearity from them to be calculated.

In several previous papers \cite{FlambaumPRA2019,AllehabiPRC2020,AllehabiPRA2021}, we examined the effect of the nuclear deformation on isotope shift in heavy elements. We  calculated  various contributions to the non-linearity of the King plots for Yb\textsuperscript{+} in Ref. \cite{AllehabiPRA2021}. An important conclusion was that  the non-linearity from the variation of the nuclear deformation in different isotopes is comparable to the observed non-linearity in Ref. \cite{CountsPRL2020}. We also found variation in the non-linearity between the results of the calculations, based on  different nuclear models, which is caused by the variation of the nuclear shape between these models.
%, in particular nuclear shape the main cause of this variation instead of the change of nuclear nuclear RMS charge radius. 
The sensitivity of the non-linearity calculations is also emphasised in Ref. \cite{MullerPRA2021}, which found that experimental uncertainties in the nuclear charge radii can lead to large variations in the non-linearity.

Nuclear polarisation has been suggested as a non-negligible contribution to the non-linearity in Ref.  \cite{FlambaumPRA2018}. In a recent paper \cite{FlambaumPRA2021}, the effect of the nuclear polarisation on spectra of atoms and ions was examined. To incorporate the effect  into  the standard atomic many-body calculations, the results of the direct calculations of the corresponding energy shifts were accurately fitted by the effective  potentials, which may be added to the Coulomb nuclear potential. 
%notably deriving effective potentials for scalar nuclear polarisation corrections in medium and heavy atoms (it was found that nuclear tensor polarisation effects are negligible). 
In particular, the  contributions to the  nuclear polarisation effects  from the giant electric-dipole (E1) nuclear resonance  and rotational nuclear (E2) transitions have been calculated, with the effective potentials derived for each contribution. 

The giant resonance contribution is enhanced by the large collective E1 nuclear matrix elements. There is a simple explanation why E2 effect is important in heavy atoms with a deformed nucleus.  In light atoms nuclear rotation is faster than electron motion, and electron   sees averaged  spherical Coulomb nuclear potential.  In heavy atoms nuclear rotation is slower  but electron moves faster. It sees  a non-spherical nucleus, and this affects  electron energy levels and wave functions. If we use perturbation theory to calculate  this effect, it is described as  a second order perturbation theory sum with the non-diagonal matrix elements of the quadruple component (E2) of the electron-nucleus electrostatic interaction, where in intermediate state both electron and  nucleus are excited. In heavy atoms typical virtual electron excitation energy is  bigger than the minimal nuclear excitation energy, which is  equal to the rotational interval.   Small nuclear excitation energy (the rotational interval) and collective nuclear matrix elements  between the rotational states  are the reasons for the enhancement of this E2 nuclear polarisation effect.  Both the nuclear E2 matrix elements and the rotational intervals depend on the isotope.  An isotopic dependence of this effect indicates an important contribution to the King plot non-linearity. 
%Further discussion on nuclear polarisation and the effective potentials is in Sec. \ref{NuclearPolarisation}, as the effective potentials can be added to FIS calculations, allowing us to calculate nuclear polarisation contributions to King plot non-linearity.

In this paper, we examine and calculate nuclear polarisation contributions to the King plot non-linearity. 
We also study the role of the relativistic effects and demonstrate that their contribution to the non-linearity is small.
We present a way of separation of the relativistic effects from other possible contributions to the non-linearities.
We consider Ca$^+$, Sr$^+$ and Yb$^+$ ions since they have already been used in the search for new interactions via non-linearities of King plot~\cite{ShiAPB2016,GorgesJPB2015,SolaroPRL2020,KnollmannPRA2019,ManovitzPRL2019,DubostPRA2014,CountsPRL2020}.

\section{King plot and contributions to  its non-linearity}\label{Kingplot}

In this work we study a contribution to the non-linearity of King plot caused by nuclear polarizability and compare it to other possible sources of the non-linearities, which exist within the standard model (SM). This includes quadratic field shift (QFS), relativistic and nuclear deformation effects.
An expression for the IS, which contains all these effects, can be written as 
\begin{eqnarray} \label{e:nu}
    &&\nu_a = F_a \delta\langle r^2 \rangle + K_a\mu^{AA'} + \\
    &&G_a^{(2)}\delta\langle r^2 \rangle^2 + G_a^{(4)}\delta\langle r^4 \rangle + \nu_a^{NP} \nonumber.
\end{eqnarray}
Here $\nu_a$ is the IS for an atomic transition $a$, $K_a$, $F_a$, $G_a^{(2)}$, and $G_a^{(4)}$ are electronic structure factors, which depend on the transition but do not depend on the nuclear isotope; $\delta\langle r^2 \rangle$, and $\delta\langle r^4 \rangle$ are changes of the nuclear moments between isotopes $A$ and $A'$, $\mu^{AA'} = 1/A-1/A'$. First line in (\ref{e:nu}) is the standard expression for the IS contaning leading contribution to the FIS and the mass shift.
Second line contains higher-order contributions to FIS which we study in present work. The term with $G_a^{(2)}$ is the QFS, the term with $G_a^{(4)}$ is due to the relativistic effects and effects of changing nuclear shape. The last term is the main focus of present work. It presents the change of the frequency of an atomic transition between two isotopes due to the difference in nuclear polarisation potentials for the two isotopes. Note that in this term we do not separate nuclear and electronic variables. This is because the nuclear polarisation potential depends on several nuclear parameters (see below) and separation of the variables leads to unwanted complications.

If we have two atomic transitions $a$ and $b$ and two corresponding equations (\ref{e:nu}), then finding $\delta\langle r^2 \rangle$ from one equation and substituting it to another leads to
%\begin{equation}\label{SM_IS_eqn}\begin{aligned}
%    \frac{\nu_b^{AA'}}{\mu^{AA'}} = \;&\frac{G_b}{F_a}\frac{\nu_a^{AA'}}{\mu^{AA'}} + \bigg(K_b - \frac{F_b}{F_a}K_a\bigg)\\ + \bigg(G_b^{(2)} - \frac{F_b}{F_a}G_a^{(2)}\bigg)\frac{\delta\langle r^2\rangle^2}{\mu^{AA'}}+ \bigg(G_b^{(4)} - \frac{F_b}{F_a}G_a^{(4)}\bigg)\frac{\delta\langle r^4\rangle}{\mu^{AA'}}.
%\end{aligned}\end{equation}
\begin{eqnarray}\label{SM_IS_eqn}
&&    \frac{\nu_b^{AA'}}{\mu^{AA'}} = \frac{G_b}{F_a}\frac{\nu_a^{AA'}}{\mu^{AA'}} + \bigg(K_b - \frac{F_b}{F_a}K_a\bigg)+\\ 
&& \bigg(G_b^{(2)} - \frac{F_b}{F_a}G_a^{(2)}\bigg)\frac{\delta\langle r^2\rangle^2}{\mu^{AA'}}+ \bigg(G_b^{(4)} - \frac{F_b}{F_a}G_a^{(4)}\bigg)\frac{\delta\langle r^4\rangle}{\mu^{AA'}} \nonumber\\
&&+ \bigg(\nu_b^{NP} - \frac{F_b}{F_a}\nu_a^{NP}\bigg)\frac{1}{\mu^{AA'}} \nonumber.
\end{eqnarray}
Here first line is the equation for the standard King plot. If we keep these terms only, points, corresponding to pairs of isotopes, are all on the same line on the $(\nu/\mu)_a$, $(\nu/\mu)_b$ plane. The terms in the second and third lines can cause the non-linearities of the King plot since they depend on isotopes. Note that each of these terms may vanish if $G_a^{(2)}/G_b^{(2)} = F_a/F_b$, or $G_a^{(4)}/G_b^{(4)} = F_a/F_b$, or $\nu_a^{NP}/\nu_b^{NP} = F_a/F_b$.

\section{Calculation of the electronic structure factors}.

In this work we calculate electronic structure factors $F$, $G^{(2)}$, $G^{(4)}$ and frequency shift $\nu^{NP}$ (see Eq.~(\ref{e:nu})). 
%We do not calculate mass shift since it does not contribute to the non-linearities of King plot. 
We consider Ca$^+$, Sr$^+$, and Yb$^+$ ions,
which all have similar electronic structure, having one external electron above closed-shell core. Therefore, we use the same approach for all three systems, the approach which we used for Yb$^+$ calculations in our previous work~\cite{AllehabiPRA2021}. In short, we perform the calculations in the $V^{N-1}$ approximation, starting from the relativistic Hartree-Fock calculations for the closed-shell core. The states of external electron are calculated in the field of the frozen core.
Core-valence correlations are included by means of the correlation potential method~\cite{Dzuba1987_CorPotMethod}. The correlation potential $\hat \Sigma$ (which is a non-local integration operator)  is calculated in the second order of the many-body perturbation theory and added to the relativistic Hartre-Fock potential. Solving Hartree-Fock equations with  added  $\hat \Sigma$ redefines the valence orbitals leading to the so-called Bruekner orbitals (BO) for the valence states. FIS is calculated as a perturbation caused by the change of the nuclear potentials between two isotopes. Random phase approximation (RPA) is used to calculate the energy shift. When the shift is caused by the change of nuclear volume and shape between two isotopes, the fitting of the RPA results leads to the determination of the $F$ and  $G^{(4)}$ electronic structure factors. The $G^{(2)}$ factor , responsible for the quadratic isotope shift, is calculated as a second-order correction to the energy (see~\cite{AllehabiPRA2021} for details). When the shift is caused by the difference in the nuclear polarisation potentials between two isotopes, the RPA calculations produce the values of $\nu^{NP}$. Calculated values of the $F$, $G^{(2)}$, $G^{(4)}$ factors for several states or transitions in Ca$^+$, Sr$^+$, and Yb$^+$ are presented in Table~\ref{t:FG}. Note that when the factors are presented for the states, the factors for transitions should be calculated as a difference between corresponding factors for the states. The accuracy of the calculations is $\sim$~1\%. However, we present many digits in the calculated values of the factors because the non-linearities are very sensitive to the ratio of the factors (see last sentence of the previous section). Since calculations for different states are done exactly the same way, we believe that the accuracy for the ratios of the electronic factors $F$, $G$ and  $\nu^{NP}$ is higher than 1\%.

\begin{table}
\caption{\label{t:FG}
Calculated electronic structure factors. Numbers in square brackets mean powers of 10.}
\begin{ruledtabular}
\begin{tabular}{ll rrr}
\multicolumn{1}{c}{Ion}&
\multicolumn{1}{c}{State or}&
\multicolumn{1}{c}{$F$} &
\multicolumn{1}{c}{$G^{(2)}$} &
\multicolumn{1}{c}{$G^{(4)}$} \\
&\multicolumn{1}{c}{transition} &
\multicolumn{1}{c}{GHz/fm$^2$}&
\multicolumn{1}{c}{GHz/fm$^4$}&
\multicolumn{1}{c}{GHz/fm$^4$}\\
\hline
Ca$^+$ & $4s_{1/2}$ &  0.26026 & -0.8358[-4] &  -0.3710[-4] \\
             & $4p_{1/2}$ & -0.02240 & -0.1525[-6] &   0.3580[-5] \\
             & $4p_{3/2}$ & -0.02262 & -0.3996[-10]&   0.2965[-5] \\
            & $3d_{3/2}$ & -0.10735 & -0.6488[-8] &   0.1569[-4] \\
            & $3d_{5/2}$ & -0.10673 & -0.6455[-8] &   0.1615[-4] \\
         	      		  	         
Sr$^+$ & $5s_{1/2}$ &  1.25903 &  -0.9380[-3] &  -0.4268[-3] \\
            & $5p_{1/2}$ & -0.07221 &  -0.7237[-5] &   0.2511[-4] \\
            & $5p_{3/2}$ & -0.08055 &   0.2619[-8] &   0.2767[-4] \\
            & $4d_{3/2}$ & -0.35372 &  -0.7741[-7] &   0.1185[-3] \\
         & $4d_{5/2}$ & -0.34346 &  -0.7529[-7] &   0.1151[-3] \\
       
Yb$^+$ & $6s-5d_{5/2}$ & -17.6035 &   0.02853    &   0.013083 \\
       & $6s-5d_{3/2}$ & -18.0028 &   0.02853    &   0.013371 \\
\end{tabular}
\end{ruledtabular}
\end{table}

\subsection{Nuclear Polarisation}\label{NuclearPolarisation}

The general effective potential for the nuclear polarisation contribution to the atomic energy shift  was introduced in Ref.~\cite{FlambaumPRA2018} to allow for calculation in multielectron atoms and ions. It may be schematically presented as 
\begin{equation}
V_L(r)=-\frac{e^2}{2} \frac{\alpha^{EL}_0}{r^{2L+2}+b^{2L+2}}\,,
\label{eff-pot}
\end{equation}
with the scalar nuclear polarizability due to the $EL$ nuclear transition from state of the total angular momentum $L$ to the ground state
\begin{equation}
    \alpha^{EL}_0 = \frac{8\pi}{2L+1} \frac{B(EL;L\to0)}{E_L}, 
\end{equation}
$b$ in (\ref{eff-pot}) is the cut-off parameter which depends on the nuclear properties, notably nuclear charge $Z$, mass number $A$, nuclear radius $R_0$, and nuclear deformation $\beta_2$. It was calculated and tabulated in ~\cite{FlambaumPRA2018}. At large distances $r>>b$ this potential  becomes an ordinary polarisation potential. Parameters of this effective potential have been fitted to reproduce directly calculated energy shifts to the accuracy $\sim$ 1 \%.

The electric dipole component of this potential  is dominated by the contribution of the  nuclear giant electric dipole resonance. The potential (\ref{eff-pot}) becomes
\begin{equation}\label{e:V1}
	V_1=-\frac{e^2}{2}\frac{ \alpha_0^{E1} }{r^4+b^4}\,,
	\quad
	\alpha_0^{E1}=\frac{ 8\pi B(E1)}{3E_{\rm GR}}\, .
\end{equation}
Here $\alpha_0^{E1}$ is the scalar electric dipole nuclear polarizability, $B(E1)$ is the giant resonance transition probability,
\begin{equation}\label{e:BE1}
B(E1)= \frac{3}{8\pi} \frac{Z(A-Z)e^2}{AE_{GR}m_p},
\end{equation}
$A$ is atomic number, $E_{GR}$ is the energy of the giant dipole resonance, $m_p$ is proton mass.

The electric quadrupole component of the  potential (\ref{eff-pot}) in deformed nuclei  is dominated by the nuclear rotational transition due to the small excitation energy $E_{rot}$ and large collective matrix element of such transition. 
\begin{equation}\label{e:V2}
V_2=-\frac{e^2}{2}\frac{\bar{\alpha}_0^{E2}}{r^6+\tilde{b}^6}\,. 
\end{equation}
Here  coefficient $\bar \alpha_0^{E2}$ is equal to 
\begin{equation}\label{alpha_0_E2}
	\bar{\alpha}_0^{E2}=\frac{8\pi}{5}\frac{B(E2)}{\bar E_{\rm rot}}\,.
\end{equation}
Note that $\bar \alpha_0^{E2}$   looks similar but not exactly equal to the  electric quadrupole scalar nuclear polarisability $\alpha_0^{E2}$. The reason is that   $\bar E_{\rm rot}= 50$ keV is a parameter, which is numerically close to the average nuclear rotational interval, but the actual dependence on the rotational energy $E_{rot}$ in a specific isotope is weak due to the domination of the electron excitation energy $E_{\rm electron}$ in the energy  denominators of the perturbation theory ($E_{\rm electron}  + E_{\rm rot}$);   $B(E2)$ is the nuclear rotational transition probability
\begin{equation}\label{e:BE2}
B(E2)= \frac{1}{5}\frac{3}{4\pi} Z^2e^2R_0^4 \beta_2^2,
\end{equation}
$R_0$ is the nuclear radius, $\beta_2$ is the parameter of the nuclear quadrupole deformation.
Note that the potential (\ref{e:V2}) vanishes for spherical nuclei.
Strong dependence on the isotope appears here due to variations of the quadrupole  deformation parameter $\beta_2$ between two isotopes, similar to the  deformation effect in the field isotope shift resulting in  the non-linearity of the King plot  \cite{AllehabiPRA2021}.

Numerical values for all parameters needed for the calculations can be found in Ref.~\cite{FlambaumPRA2018}. 

\section{Results}

\subsection{Non-linearities of King plot}

To calculate non-linearities of King plot we calculate IS for a pair of transition using equation (\ref{e:nu}) and calculate function (\ref{SM_IS_eqn}) which corresponds to the King plot. Mass shift is not included in the calculations since it does not contribute to the non-linearities. It just moves the King plot on the vertical scale without affecting its shape. On the next step we fit the function (\ref{SM_IS_eqn}) by a straight line on the $\nu_a/\mu$, $\nu_b/\mu$ plane. The non-linearities are defined as relative deviation $r$ of the calculated FIS from the straight line,
\begin{equation}\label{e:NL}
%r^{AA'} = \frac{\nu_b^{AA'}/\mu^{AA'} - \nu_{sl}^{AA'}}{\nu_b^{AA'}/\mu^{AA'}}.
r^{AA'} = \frac{\tilde \nu_b^{AA'} - \nu_{sl}^{AA'}}{\tilde \nu_b^{AA'}},
\end{equation}
where $\tilde \nu \equiv \nu/\mu$ is given by formula (\ref{SM_IS_eqn}), $\nu_{sl}$ is the result of fitting by straight line.

To calculate FIS using (\ref{e:nu}) we need to know electronic structure factors $F$, $G^{(2)}$, $G^{(4)}$, nuclear factors $\delta \langle r^2 \rangle$, $\delta \langle r^4 \rangle$, and nuclear parameters for the nuclear polarisation potentials $V_1$ and $V_2$ (see Eqs. (\ref{e:V1},\ref{e:V2})).
Calculation of the electronic structure factors was considered in previous section and described in more details in our earlier paper on Yb$^+$~\cite{AllehabiPRA2021}. 
For the nuclear structure factors $\delta \langle r^2 \rangle$, $\delta \langle r^4 \rangle$ we use the results of nuclear calculations performed by means of Hartree-Fock-Bogolubov method with the covariant density functional called DD-PC1~\cite{DDPC1}. For Yb$^+$ we also use the FIT1 nuclear model~\cite{AllehabiPRA2021} in which nuclear parameters were chosen to fit the observed non-linearities of the King plot. Note that nuclear calculations do not give the value of the  $\delta \langle r^4 \rangle$ directly. Instead, they give the value of the nuclear quadrupole deformation parameter $\beta_2$, which can be used to calculate the nuclear density averaged over nuclear rotation. Using this density we  calculate corresponding  $\delta \langle r^4 \rangle$ (see Ref.~\cite{AllehabiPRA2021} for more details). For the nuclear polarisation potentials $V_1$ and $V_2$ we use nuclear parameters presented in Ref. \cite{FlambaumPRA2018}.

To separate relative contributions to the non-linearities from different effects, such as QFS, nuclear shape and nuclear polarisation, we remove in turn all but one term from the second and third line of Eq. (\ref{SM_IS_eqn}), do the fitting by the straight line and calculate the non-linearities using Eq. (\ref{e:NL}).
The results are presented in Table~\ref{t:NL}. We study the non-linearities for several pairs of transitions, following experimental publications. E.g. the $4s - 4p_{1/2,3/2}$ pair of transitions in Ca$^+$ ion was studied in Refs.~\cite{ShiAPB2016,GorgesJPB2015}, the $4s-3d_{3/2,5/2}$ pair of transitions in Ca$^+$ ion was studied in Ref.~\cite{SolaroPRL2020}, the $5s-5p_{1/2}$, $4d_{3/2}-5p_{1/2}$ pair of transitions in Sr$^+$ ion was studied in Ref.~\cite{DubostPRA2014}, and the $6s-5d_{3/2,5/2}$ pair of transitions in Yb$^+$ ion was studied in Ref.~\cite{CountsPRL2020}.

It turns out that the results are sensitive to the details of the calculations. In particular, they are sensitive to the values of the nuclear parameters. Therefore, the uncertainties are close to the values of the the calculated non-linearities. However, the order of magnitude for each source of the non-linearities is established sufficiently reliably. We see that for all considered ions the dominating contribution to the non-linearity of King plot comes from the QFS and nuclear shape effects (terms associated with the $G^{(2)}$ and $G^{(4)}$ coefficients in (\ref{e:nu}) and (\ref{SM_IS_eqn})), while the contribution from nuclear polarisation is significantly smaller.
%We see that for Ca$^+$ and Sr$^+$ the dominating source of non-linearity is QFS (the term with $G^{(2)}$). 
In Yb$^+$ the dominating source of the  non-linearity is nuclear deformation (associated with the $G^{(4)}$ coefficient) in agreement with our earlier study~\cite{AllehabiPRA2021}. 
All considered isotopes of Ca and Sr have spherical nuclei. Therefore, the $V_2$ nuclear polarisation potential is zero for them and there is no corresponding contribution to the non-linearities. The calculated non-linearities in the FIT1 nuclear model for Yb$^+$ are in good agreement with experiment, which is the result of the fitting by the nuclear deformation parameters (see Ref.~\cite{AllehabiPRA2021} for details). 
The calculated non-linearities in Ca$^+$ and Sr$^+$ are consistent with the limits found in experiments \cite{ShiAPB2016,GorgesJPB2015,DubostPRA2014}.
%The contributions from $V_1$  nuclear polarisation potential is two to three orders of magnitude smaller than the dominating contribution. In Yb$^+$ the contribution from $V_2$ is about two orders of magnitude smaller that the dominating contribution (nuclear deformation) while the contribution from $V_1$ is about three orders  of magnitude smaller.

It might be useful to know the values of absolute contributions of the nuclear polarisation to the IS. According to the calculations, the contribution from the $V_1$ potential is $\sim$ 100 to 400~Hz for Ca$^+$, $\sim$ 1.0 to 1.6~kHz in Sr$^+$, and $\sim$ 4~kHz in Yb$^+$. The contribution from $V_2$ to IS in Yb$^+$ is $\sim \pm$~100~Hz. 

%We study 
%Ca+   V1   ~100 - 400 Hz
%Sr+   V1   ~1.0 - 1.6 kHz
%Yb+   V1   ~4.0       kH
%      V2   ~ +/- 100 Hz

We see that nuclear polarization is not the dominating contribution to the non-linearity of King plot. However, it is sufficiently large to affect the analysis of the non-linearities in terms of the  manifestation of new interactions.

 The accuracy of the calculation of all considered sources of the non-linearities (QFS, nuclear deformation and nuclear polarisation) is insufficient to subtract  them from the observed non-linearities and use the residual non-linearities to put the limits on a new interaction. 
Therefore, the limits on new boson may be determined by the uncertainty of  SM contributions to the non-linearity rather than experimental errors. Order of magnitude estimates  for this situation have been presented in the Table V of Ref.  \cite{FlambaumPRA2018}.  SM contributions to the non-linearity increase with the nuclear charge $Z$.  If new boson Compton wave length is comparable or bigger than Bohr radius $a_B$, the new boson contribution  to the non-linearity does not increase with  $Z$ and one could conclude  that the best limits on the new boson  may be obtained in light atoms with spherical nucleus like Ca, where the field  shift contributions to the non-linearities  are small. However, a proper estimate of the non-linearities from the quadratic mass shift is required here, and this may become another theoretical problem. 
Atoms with medium spherical nuclei like Sr have much smaller quadratic mass shift than Ca. Also, if the Compton wave length of new boson is smaller than Bohr radius, the contribution of the new boson to the  non-linearities  increases with $Z$. This gives another advantage to Sr in comparison with Ca.  

\begin{table}
\caption{\label{t:NL}
Non-linearities of King plot coming from different sources. Numbers in square brackets mean powers of 10.}
\begin{ruledtabular}
\begin{tabular}{c rrrrr}
\multicolumn{1}{c}{Pair of}&
\multicolumn{1}{c}{$G^{(2)}$} &
\multicolumn{1}{c}{$G^{(4)}$} &
\multicolumn{1}{c}{$V_1$} &
\multicolumn{1}{c}{$V_2$} &
\multicolumn{1}{c}{Expt.} \\
\multicolumn{1}{c}{isotopes}&&&&&
\multicolumn{1}{c}{\cite{CountsPRL2020}}\\
\hline
%&&&&&\\
\multicolumn{6}{c}{Ca$^+$, DD-PC1, $4s-4p_{1/2}$ and $4s-4p_{3/2}$} \\
%&&&&&\\
%40  -  42 &  3.6[-6] &  3.0[-9] &   4.1[-9] &           & \\   
%42  -  44 & -7.2[-7] & -6.3[-8] &  -8.2[-9] &	        & \\   
%44  -  46 & -5.6[-8] &  6.3[-10]& -6.6[-10] &	        & \\   
%46  -  48 &  5.5[-6] &  3.7[-9] &   6.3[-9] &	        & \\   
%Ca$^+$, DD-PC1, 4s-4p 1/2, 4s - 4p 3/2
40  -  42 & -1.7[-9] &  2.2[-8] &  -3.2[-9] &           & \\ 
42  -  44 &  3.3[-9] & -4.6[-8] &   6.4[-9] &	        & \\  
44  -  46 &  2.6[-10]&  4.6[-9] &   5.1[-10] &	        & \\  
46  -  48 & -2.5[-9] &  2.7[-8] &  -4.9[-9] &	        & \\
%&&&&&\\	      	   	      	         	     	         
\multicolumn{6}{c}{Ca$^+$, DD-PC1, $4s-3d_{5/2}$ and $4s-3d_{3/2}$} \\
%Ca$^+$, DD-PC1, 4s-3d 5/2, 3s - 3d 3/2
40  -  42 &  2.1[-9] &  1.4[-8] &  -4.3[-10] &           & \\ 
42  -  44 & -4.1[-9] & -3.0[-8] &   8.6[-10] &	        & \\  
44  -  46 & -3.1[-10]&  3.0[-9] &   6.7[-11] &	        & \\  
46  -  48 &  3.2[-9] &  1.7[-8] &  -6.5[-10] &	        & \\
%\multicolumn{6}{c}{Sr$^+$, DD-PC1, $5s-5p_{1/2}$ and $5s-4d_{3/2}$} \\
%&&&&&\\
%84  -  86 &  5.5[-6] &  3.8[-7] &  6.3[-10] &	        & \\   
%86  -  88 &  6.2[-6] &  1.3[-7] &   3.2[-9] & 	        & \\   
%88  -  90 & -2.2[-6] & -9.8[-8] & -6.7[-10] &	        & \\   
%90  -  92 & -6.6[-5] & -2.9[-6] &  -2.1[-8] & 	        & \\   
%&&&&&\\  	      	         	     	         
\multicolumn{6}{c}{Sr$^+$, DD-PC1, $5s-5p_{1/2}$ and $5p_{1/2}-4d_{3/2}$} \\
%&&&&&\\
84  -  86 &  3.1[-5] &  2.2[-6] &  3.6[-9] &	        & \\   
86  -  88 &  3.6[-5] &  7.7[-7] &  1.8[-8] & 	        & \\   
88  -  90 & -1.2[-5] & -5.6[-7] & -3.8[-9] &	        & \\   
90  -  92 & -3.7[-4] & -1.7[-5] & -1.2[-7] & 	        & \\   
%&&&&&\\
\multicolumn{6}{c}{Yb$^+$, DD-PC1, $6s-5d_{3/2}$ and $6s-5d_{5/2}$} \\
%&&&&&\\
168 - 170 & -4.5[-8] &  1.2[-8] & -1.6[-10] & -3.1[-9]  & -1.9[-7] \\
170 - 172 & -2.9[-7] & -2.4[-7] &  -1.3[-9] & -3.3[-8]  &  2.7[-7] \\
172 - 174 & -8.8[-8] &  4.0[-7] &  1.0[-11] &  7.8[-9]  & -4.9[-7] \\
174 - 176 &  4.0[-7] & -1.6[-7] &   1.5[-9] &  2.6[-8]  &  4.1[-7] \\
%&&&&&\\	      	   	      	         	     	         
\multicolumn{6}{c}{Yb$^+$, FIT1, $6s-5d_{3/2}$ and $6s-5d_{5/2}$} \\
%&&&&&\\
168 - 170 & -6.4[-8] & -1.6[-7] & -4.2[-11] & -2.5[-9]  & -1.9[-7] \\
170 - 172 &  7.9[-8] &  2.2[-7] &  1.0[-10] &  3.5[-9]  &  2.7[-7] \\
172 - 174 & -2.3[-8] & -4.3[-7] & -6.2[-10] & -5.9[-9]  & -4.9[-7] \\
174 - 176 &  6.8[-9] &  3.6[-7] &  5.6[-10] &  4.8[-9]  &  4.1[-7] \\
\end{tabular}
\end{ruledtabular}
\end{table}

\subsection{The Role of Relativistic Effects}

In non-relativistic case FIS is given by
\begin{equation}\label{e:nrFIS}
\nu^{\rm FIS} = F\delta \langle r^2 \rangle.
\end{equation}
This formula is widely used to extract the change of nuclear charge radius from the IS measurements. 
It is even used for this purpose for heavy elements up to nobelium ($Z$=102) \cite{No-IS-exp}.
However, in heavy atoms the use of formula (\ref{e:nrFIS}) should be reconsidered due to the role of relativistic effects.
They manifest themselves, in particular, in the dependence of the electronic structure factor $F$ in (\ref{e:nrFIS}) on the nuclear radius.
This dependence can be neglected for neighbouring isotopes for the extraction of $\delta \langle r^2 \rangle$ (which was the case for nobelium).
But it is not the case when the non-linearities of the King plot are considered. The separation of the nuclear and electronic variables is a necessary condition for the King plot to be linear. Given that the non-linearities can be very small and at least four isotopes are needed to study them, the drift of the coefficient  $F$ before $\delta \langle r^2 \rangle$ can no longer be neglected.
It was suggested in Ref.~\cite{FlambaumPRA2018} to use a different formula
\begin{equation}\label{e:rFIS}
\nu^{\rm FIS} = F'\delta \langle r^{2\gamma} \rangle,
\end{equation}
where $\gamma=\sqrt{1-(Z\alpha)^2}$, $\alpha =1/137.036$ is the fine structure constant.
In this formula the electronic structure factor $F'$ practically does not depend on the  nuclear isotope. It was obtained by the relativistic consideration of a model problem in which the nucleus was presented as a uniformly charged ball and approximate analytical expressions for the wave functions for the $s$ and $p_{1/2}$ states inside the nucleus were used. A more sophisticated formula was suggested in Ref.~\cite{MullerPRA2021} using a similar approach.

In this section we study the problem numerically, using a more realistic nuclear model with the Fermi distribution of nuclear charge and numerical wave functions obtained in the relativistic Hartree-Fock calculations or in the numerical calculations for the hydrogen -like ions.
FIS is calculated as an expectation value of the change of the nuclear potential between two nuclear isotopes. The results are presented in the form, similar to (\ref{e:rFIS}),
\begin{equation}\label{e:lambda}
\nu^{\rm FIS}_a = \langle a| \delta V_N|a\rangle = F_a''\delta \langle r^{\lambda} \rangle.
\end{equation}
The values of $\lambda$ are found from the condition that the electronic structure factor $F''$ does not depend on the nuclear isotope. The results are presented in Table~\ref{t:lambda} for $s_{1/2}$ and $p_{1/2}$ wave functions of Ca$^+$, Sr$^+$, Yb$^+$, Fr and E120$^+$. It turns out that the values of $\lambda$ do not depend on which kind of single-electron orbitals are used, relativistic Hartree-Fock orbitals, Bruekner orbitals (the orbitals which include core-valence correlations~\cite{Dzuba1987_CorPotMethod}), or the orbitals obtained for the hydrogen - like systems. Note that the values of $\lambda$ for  $s_{1/2}$ and $p_{1/2}$ states are very close. This means that in the approximation, when this difference is neglected, there is no contribution to the non-linearities of King plot. This corresponds to the case $F_a/F_b = G^{(4)}_a/G^{(4)}_b$ in (\ref{SM_IS_eqn}).

The values of the power $\lambda$ are close to $2\gamma$ for light atoms only. For heavy atoms the difference is significant. This is due to the difference in the analytical and numerical considerations. Only leading in $(Z\alpha)^2$ term was included in the expansion of the electronic density inside the nucleus in Ref.  \cite{FlambaumPRA2018}. This works well for small $Z$ (accurate to $\leq$ 1\% for $Z \leq 50$), but becomes increasingly inaccurate for large $Z$.
Numerical results show that even for superheavy atoms the values of $\lambda$ are closer to 2 than to $2\gamma$. This justifies the use of (\ref{e:nrFIS}) for the extraction of the change of nuclear radius between neighbouring isotopes of heavy atoms.
The value of $\lambda$ for any nuclear charge $Z$ can be interpolated with the accuracy $\sim$~1\% by the formula
\begin{equation}\label{e:lambdafit}
\lambda(Z)=2 - 3.1968\times 10^{-4} Z - 2.0632\times 10^{-5}Z^2.
%      g2=2.-0.319677E-03*iz-0.206316E-04*iz**2
\end{equation}
This formula is the result of fitting of the data from Table~\ref{t:lambda}.

Table \ref{t:lambda} shows results assuming  the spherical nuclear charge distribution described by the standard Fermi-type  formula. When nuclear deformation is taken into account, formula (\ref{e:lambda}) becomes less accurate.  The way of dealing with the nuclear deformation was considered in our earlier paper~\cite{AllehabiPRA2021}.

\begin{table}
\caption{\label{t:lambda}
Parameters $\lambda$ (formula (\ref{e:lambda})) for the $s_{1/2}$ and $p_{1/2}$ functions of different atomic systems.
$R_N$ is nuclear radius (parameter of Fermi charge distribution) used in the calculations.}
\begin{ruledtabular}
\begin{tabular}{rrrrr}
\multicolumn{1}{c}{$Z$}&
\multicolumn{1}{c}{$R_N$(fm)}&
\multicolumn{1}{c}{$\lambda_{s_{1/2}}$}&
\multicolumn{1}{c}{$\lambda_{p_{1/2}}$}&
\multicolumn{1}{c}{$2\gamma$}\\
\hline
 20 &  3.719 &   1.967 &   1.968 &   1.9786 \\
 38 &  4.867 &   1.963 &   1.966 &   1.9216 \\
 70 &  6.277 &   1.881 &   1.887 &   1.7194 \\
 87 &  6.834 &   1.818 &   1.826 &   1.5452 \\
120 &  7.360 &   1.662 &   1.673 &   0.9658 \\
\end{tabular}
\end{ruledtabular}
\end{table}

\section*{Conclusion}

In this paper we calculated the contribution of the nuclear polarisation into isotope shift  and non-linearities of the  King plot for Ca$^+$, Sr$^+$ and Yb$^+$ ions.
We also studied the role of the relativistic effects and found that their contribution to the non-linearities is small if we assume spherical distribution of the nuclear charge. A formula was found which allows us  to isolate relativistic effects from other sources of the non-linearities. We argue that the accuracy of the calculations of the non-linearities of the King plot, which come from  the field shift, is insufficient to isolate them from possible contribution of a new interaction.  
Therefore, the limits on new boson may be determined by the uncertainty of  SM contributions to the non-linearity rather than experimental errors. An important source of SM non-linearities is variation of the nuclear deformation from isotope to isotope.  Therefore, the best choices for the new boson search are among atoms with medium mass spherical nuclei, like Sr and Ca, where both the field shift and quadratic mass shift contributions to the non-linearities are suppressed in comparison with deformed heavy nuclei and very light nuclei.    

\section*{Acknowledgements}
The work work supported in part by the Australian Research Council.

%\bibliographystyle{apsrev4-2}
%\bibliographystyle{apsrev}
%\bibliography{biblio,dzuba}
%\bibliography{dzuba}

\begin{thebibliography}{23}
\expandafter\ifx\csname natexlab\endcsname\relax\def\natexlab#1{#1}\fi
\expandafter\ifx\csname bibnamefont\endcsname\relax
  \def\bibnamefont#1{#1}\fi
\expandafter\ifx\csname bibfnamefont\endcsname\relax
  \def\bibfnamefont#1{#1}\fi
\expandafter\ifx\csname citenamefont\endcsname\relax
  \def\citenamefont#1{#1}\fi
\expandafter\ifx\csname url\endcsname\relax
  \def\url#1{\texttt{#1}}\fi
\expandafter\ifx\csname urlprefix\endcsname\relax\def\urlprefix{URL }\fi
\providecommand{\bibinfo}[2]{#2}
\providecommand{\eprint}[2][]{\url{#2}}

\bibitem[{\citenamefont{Counts et~al.}(2020)\citenamefont{Counts, Hur,
  Aude~Craik, Jeon, Leung, Berengut, Geddes, Kawasaki, Jhe, and
  Vuleti\ifmmode~\acute{c}\else \'{c}\fi{}}}]{CountsPRL2020}
\bibinfo{author}{\bibfnamefont{I.}~\bibnamefont{Counts}},
  \bibinfo{author}{\bibfnamefont{J.}~\bibnamefont{Hur}},
  \bibinfo{author}{\bibfnamefont{D.~P.~L.} \bibnamefont{Aude~Craik}},
  \bibinfo{author}{\bibfnamefont{H.}~\bibnamefont{Jeon}},
  \bibinfo{author}{\bibfnamefont{C.}~\bibnamefont{Leung}},
  \bibinfo{author}{\bibfnamefont{J.~C.} \bibnamefont{Berengut}},
  \bibinfo{author}{\bibfnamefont{A.}~\bibnamefont{Geddes}},
  \bibinfo{author}{\bibfnamefont{A.}~\bibnamefont{Kawasaki}},
  \bibinfo{author}{\bibfnamefont{W.}~\bibnamefont{Jhe}}, \bibnamefont{and}
  \bibinfo{author}{\bibfnamefont{V.}~\bibnamefont{Vuleti\ifmmode~\acute{c}\else
  \'{c}\fi{}}}, \bibinfo{journal}{Phys. Rev. Lett.}
  \textbf{\bibinfo{volume}{125}}, \bibinfo{pages}{123002}
  (\bibinfo{year}{2020}),
  \urlprefix\url{https://link.aps.org/doi/10.1103/PhysRevLett.125.123002}.

\bibitem[{\citenamefont{Knollmann et~al.}(2019)\citenamefont{Knollmann, Patel,
  and Doret}}]{KnollmannPRA2019}
\bibinfo{author}{\bibfnamefont{F.~W.} \bibnamefont{Knollmann}},
  \bibinfo{author}{\bibfnamefont{A.~N.} \bibnamefont{Patel}}, \bibnamefont{and}
  \bibinfo{author}{\bibfnamefont{S.~C.} \bibnamefont{Doret}},
  \bibinfo{journal}{Phys. Rev. A} \textbf{\bibinfo{volume}{100}},
  \bibinfo{pages}{022514} (\bibinfo{year}{2019}),
  \urlprefix\url{https://link.aps.org/doi/10.1103/PhysRevA.100.022514}.

\bibitem[{\citenamefont{Miyake et~al.}(2019)\citenamefont{Miyake, Pisenti,
  Elgee, Sitaram, and Campbell}}]{MiyakePRR2019}
\bibinfo{author}{\bibfnamefont{H.}~\bibnamefont{Miyake}},
  \bibinfo{author}{\bibfnamefont{N.~C.} \bibnamefont{Pisenti}},
  \bibinfo{author}{\bibfnamefont{P.~K.} \bibnamefont{Elgee}},
  \bibinfo{author}{\bibfnamefont{A.}~\bibnamefont{Sitaram}}, \bibnamefont{and}
  \bibinfo{author}{\bibfnamefont{G.~K.} \bibnamefont{Campbell}},
  \bibinfo{journal}{Phys. Rev. Research} \textbf{\bibinfo{volume}{1}},
  \bibinfo{pages}{033113} (\bibinfo{year}{2019}),
  \urlprefix\url{https://link.aps.org/doi/10.1103/PhysRevResearch.1.033113}.

\bibitem[{\citenamefont{Manovitz et~al.}(2019)\citenamefont{Manovitz, Shaniv,
  Shapira, Ozeri, and Akerman}}]{ManovitzPRL2019}
\bibinfo{author}{\bibfnamefont{T.}~\bibnamefont{Manovitz}},
  \bibinfo{author}{\bibfnamefont{R.}~\bibnamefont{Shaniv}},
  \bibinfo{author}{\bibfnamefont{Y.}~\bibnamefont{Shapira}},
  \bibinfo{author}{\bibfnamefont{R.}~\bibnamefont{Ozeri}}, \bibnamefont{and}
  \bibinfo{author}{\bibfnamefont{N.}~\bibnamefont{Akerman}},
  \bibinfo{journal}{Phys. Rev. Lett.} \textbf{\bibinfo{volume}{123}},
  \bibinfo{pages}{203001} (\bibinfo{year}{2019}),
  \urlprefix\url{https://link.aps.org/doi/10.1103/PhysRevLett.123.203001}.

\bibitem[{\citenamefont{Figueroa et~al.}(2021)\citenamefont{Figueroa, Budker,
  Antypas, Berengut, Dzuba, and Flambaum}}]{Yb+Yb}
\bibinfo{author}{\bibfnamefont{N.~L.} \bibnamefont{Figueroa}},
  \bibinfo{author}{\bibfnamefont{D.}~\bibnamefont{Budker}},
  \bibinfo{author}{\bibfnamefont{D.}~\bibnamefont{Antypas}},
  \bibinfo{author}{\bibfnamefont{J.~C.} \bibnamefont{Berengut}},
  \bibinfo{author}{\bibfnamefont{V.~A.} \bibnamefont{Dzuba}}, \bibnamefont{and}
  \bibinfo{author}{\bibfnamefont{V.~V.} \bibnamefont{Flambaum}}
  (\bibinfo{year}{2021}), \eprint{2111.01429}.

\bibitem[{\citenamefont{King}(1963)}]{King1963}
\bibinfo{author}{\bibfnamefont{W.~H.} \bibnamefont{King}}, \bibinfo{journal}{J.
  Opt. Soc. Am.} \textbf{\bibinfo{volume}{\textbf{53}}}, \bibinfo{pages}{638}
  (\bibinfo{year}{1963}).

\bibitem[{\citenamefont{Delaunay et~al.}(2017)\citenamefont{Delaunay, Ozeri,
  Perez, and Soreq}}]{DelaunayPRD2017}
\bibinfo{author}{\bibfnamefont{C.}~\bibnamefont{Delaunay}},
  \bibinfo{author}{\bibfnamefont{R.}~\bibnamefont{Ozeri}},
  \bibinfo{author}{\bibfnamefont{G.}~\bibnamefont{Perez}}, \bibnamefont{and}
  \bibinfo{author}{\bibfnamefont{Y.}~\bibnamefont{Soreq}},
  \bibinfo{journal}{Phys. Rev. D} \textbf{\bibinfo{volume}{96}},
  \bibinfo{pages}{093001} (\bibinfo{year}{2017}),
  \urlprefix\url{https://link.aps.org/doi/10.1103/PhysRevD.96.093001}.

\bibitem[{\citenamefont{Berengut et~al.}(2018)\citenamefont{Berengut, Budker,
  Delaunay, Flambaum, Frugiuele, Fuchs, Grojean, Harnik, Ozeri, Perez
  et~al.}}]{BerengutPRL2018}
\bibinfo{author}{\bibfnamefont{J.~C.} \bibnamefont{Berengut}},
  \bibinfo{author}{\bibfnamefont{D.}~\bibnamefont{Budker}},
  \bibinfo{author}{\bibfnamefont{C.}~\bibnamefont{Delaunay}},
  \bibinfo{author}{\bibfnamefont{V.~V.} \bibnamefont{Flambaum}},
  \bibinfo{author}{\bibfnamefont{C.}~\bibnamefont{Frugiuele}},
  \bibinfo{author}{\bibfnamefont{E.}~\bibnamefont{Fuchs}},
  \bibinfo{author}{\bibfnamefont{C.}~\bibnamefont{Grojean}},
  \bibinfo{author}{\bibfnamefont{R.}~\bibnamefont{Harnik}},
  \bibinfo{author}{\bibfnamefont{R.}~\bibnamefont{Ozeri}},
  \bibinfo{author}{\bibfnamefont{G.}~\bibnamefont{Perez}},
  \bibnamefont{et~al.}, \bibinfo{journal}{Phys. Rev. Lett.}
  \textbf{\bibinfo{volume}{120}}, \bibinfo{pages}{091801}
  (\bibinfo{year}{2018}),
  \urlprefix\url{https://link.aps.org/doi/10.1103/PhysRevLett.120.091801}.

\bibitem[{\citenamefont{Flambaum et~al.}(2018)\citenamefont{Flambaum, Geddes,
  and Viatkina}}]{FlambaumPRA2018}
\bibinfo{author}{\bibfnamefont{V.~V.} \bibnamefont{Flambaum}},
  \bibinfo{author}{\bibfnamefont{A.~J.} \bibnamefont{Geddes}},
  \bibnamefont{and} \bibinfo{author}{\bibfnamefont{A.~V.}
  \bibnamefont{Viatkina}}, \bibinfo{journal}{Phys. Rev. A}
  \textbf{\bibinfo{volume}{97}}, \bibinfo{pages}{032510}
  (\bibinfo{year}{2018}),
  \urlprefix\url{https://link.aps.org/doi/10.1103/PhysRevA.97.032510}.

\bibitem[{\citenamefont{Frugiuele et~al.}(2017)\citenamefont{Frugiuele, Fuchs,
  Perez, and Schlaffer}}]{FrugiuelePRD2017}
\bibinfo{author}{\bibfnamefont{C.}~\bibnamefont{Frugiuele}},
  \bibinfo{author}{\bibfnamefont{E.}~\bibnamefont{Fuchs}},
  \bibinfo{author}{\bibfnamefont{G.}~\bibnamefont{Perez}}, \bibnamefont{and}
  \bibinfo{author}{\bibfnamefont{M.}~\bibnamefont{Schlaffer}},
  \bibinfo{journal}{Phys. Rev. D} \textbf{\bibinfo{volume}{96}},
  \bibinfo{pages}{015011} (\bibinfo{year}{2017}),
  \urlprefix\url{https://link.aps.org/doi/10.1103/PhysRevD.96.015011}.

\bibitem[{\citenamefont{Rehbehn et~al.}(2021)\citenamefont{Rehbehn, Rosner,
  Bekker, Berengut, Schmidt, King, Micke, Gu, M\"uller, Surzhykov
  et~al.}}]{RehbehnPRA2021}
\bibinfo{author}{\bibfnamefont{N.-H.} \bibnamefont{Rehbehn}},
  \bibinfo{author}{\bibfnamefont{M.~K.} \bibnamefont{Rosner}},
  \bibinfo{author}{\bibfnamefont{H.}~\bibnamefont{Bekker}},
  \bibinfo{author}{\bibfnamefont{J.~C.} \bibnamefont{Berengut}},
  \bibinfo{author}{\bibfnamefont{P.~O.} \bibnamefont{Schmidt}},
  \bibinfo{author}{\bibfnamefont{S.~A.} \bibnamefont{King}},
  \bibinfo{author}{\bibfnamefont{P.}~\bibnamefont{Micke}},
  \bibinfo{author}{\bibfnamefont{M.~F.} \bibnamefont{Gu}},
  \bibinfo{author}{\bibfnamefont{R.}~\bibnamefont{M\"uller}},
  \bibinfo{author}{\bibfnamefont{A.}~\bibnamefont{Surzhykov}},
  \bibnamefont{et~al.}, \bibinfo{journal}{Phys. Rev. A}
  \textbf{\bibinfo{volume}{103}}, \bibinfo{pages}{L040801}
  (\bibinfo{year}{2021}),
  \urlprefix\url{https://link.aps.org/doi/10.1103/PhysRevA.103.L040801}.

\bibitem[{\citenamefont{Flambaum and Dzuba}(2019)}]{FlambaumPRA2019}
\bibinfo{author}{\bibfnamefont{V.~V.} \bibnamefont{Flambaum}} \bibnamefont{and}
  \bibinfo{author}{\bibfnamefont{V.~A.} \bibnamefont{Dzuba}},
  \bibinfo{journal}{Phys. Rev. A} \textbf{\bibinfo{volume}{100}},
  \bibinfo{pages}{032511} (\bibinfo{year}{2019}),
  \urlprefix\url{https://link.aps.org/doi/10.1103/PhysRevA.100.032511}.

\bibitem[{\citenamefont{Allehabi et~al.}(2020)\citenamefont{Allehabi, Dzuba,
  Flambaum, Afanasjev, and Agbemava}}]{AllehabiPRC2020}
\bibinfo{author}{\bibfnamefont{S.~O.} \bibnamefont{Allehabi}},
  \bibinfo{author}{\bibfnamefont{V.~A.} \bibnamefont{Dzuba}},
  \bibinfo{author}{\bibfnamefont{V.~V.} \bibnamefont{Flambaum}},
  \bibinfo{author}{\bibfnamefont{A.~V.} \bibnamefont{Afanasjev}},
  \bibnamefont{and} \bibinfo{author}{\bibfnamefont{S.~E.}
  \bibnamefont{Agbemava}}, \bibinfo{journal}{Physical Review C}
  \textbf{\bibinfo{volume}{102}} (\bibinfo{year}{2020}), ISSN
  \bibinfo{issn}{2469-9993},
  \urlprefix\url{http://dx.doi.org/10.1103/PhysRevC.102.024326}.

\bibitem[{\citenamefont{Allehabi et~al.}(2021)\citenamefont{Allehabi, Dzuba,
  Flambaum, and Afanasjev}}]{AllehabiPRA2021}
\bibinfo{author}{\bibfnamefont{S.~O.} \bibnamefont{Allehabi}},
  \bibinfo{author}{\bibfnamefont{V.~A.} \bibnamefont{Dzuba}},
  \bibinfo{author}{\bibfnamefont{V.~V.} \bibnamefont{Flambaum}},
  \bibnamefont{and} \bibinfo{author}{\bibfnamefont{A.~V.}
  \bibnamefont{Afanasjev}}, \bibinfo{journal}{Phys. Rev. A}
  \textbf{\bibinfo{volume}{103}}, \bibinfo{pages}{L030801}
  (\bibinfo{year}{2021}),
  \urlprefix\url{https://link.aps.org/doi/10.1103/PhysRevA.103.L030801}.

\bibitem[{\citenamefont{Müller et~al.}(2021)\citenamefont{Müller, Yerokhin,
  Artemyev, and Surzhykov}}]{MullerPRA2021}
\bibinfo{author}{\bibfnamefont{R.~A.} \bibnamefont{Müller}},
  \bibinfo{author}{\bibfnamefont{V.~A.} \bibnamefont{Yerokhin}},
  \bibinfo{author}{\bibfnamefont{A.~N.} \bibnamefont{Artemyev}},
  \bibnamefont{and}
  \bibinfo{author}{\bibfnamefont{A.}~\bibnamefont{Surzhykov}},
  \bibinfo{journal}{Physical Review A} \textbf{\bibinfo{volume}{104}}
  (\bibinfo{year}{2021}), ISSN \bibinfo{issn}{2469-9934},
  \urlprefix\url{http://dx.doi.org/10.1103/PhysRevA.104.L020802}.

\bibitem[{\citenamefont{Flambaum et~al.}(2021)\citenamefont{Flambaum, Samsonov,
  Tan, and Viatkina}}]{FlambaumPRA2021}
\bibinfo{author}{\bibfnamefont{V.~V.} \bibnamefont{Flambaum}},
  \bibinfo{author}{\bibfnamefont{I.~B.} \bibnamefont{Samsonov}},
  \bibinfo{author}{\bibfnamefont{H.~B.~T.} \bibnamefont{Tan}},
  \bibnamefont{and} \bibinfo{author}{\bibfnamefont{A.~V.}
  \bibnamefont{Viatkina}}, \bibinfo{journal}{Phys. Rev. A}
  \textbf{\bibinfo{volume}{103}}, \bibinfo{pages}{032811}
  (\bibinfo{year}{2021}),
  \urlprefix\url{https://link.aps.org/doi/10.1103/PhysRevA.103.032811}.

\bibitem[{\citenamefont{Shi et~al.}(2016)\citenamefont{Shi, Gebert, Gorges,
  Kaufmann, Nörtershäuser, Sahoo, Surzhykov, Yerokhin, Berengut, Wolf
  et~al.}}]{ShiAPB2016}
\bibinfo{author}{\bibfnamefont{C.}~\bibnamefont{Shi}},
  \bibinfo{author}{\bibfnamefont{F.}~\bibnamefont{Gebert}},
  \bibinfo{author}{\bibfnamefont{C.}~\bibnamefont{Gorges}},
  \bibinfo{author}{\bibfnamefont{S.}~\bibnamefont{Kaufmann}},
  \bibinfo{author}{\bibfnamefont{W.}~\bibnamefont{Nörtershäuser}},
  \bibinfo{author}{\bibfnamefont{B.~K.} \bibnamefont{Sahoo}},
  \bibinfo{author}{\bibfnamefont{A.}~\bibnamefont{Surzhykov}},
  \bibinfo{author}{\bibfnamefont{V.~A.} \bibnamefont{Yerokhin}},
  \bibinfo{author}{\bibfnamefont{J.~C.} \bibnamefont{Berengut}},
  \bibinfo{author}{\bibfnamefont{F.}~\bibnamefont{Wolf}}, \bibnamefont{et~al.},
  \bibinfo{journal}{Applied Physics B} \textbf{\bibinfo{volume}{123}},
  \bibinfo{pages}{2} (\bibinfo{year}{2016}), ISSN \bibinfo{issn}{1432-0649},
  \urlprefix\url{http://dx.doi.org/10.1007/s00340-016-6572-z}.

\bibitem[{\citenamefont{Gorges et~al.}(2015)\citenamefont{Gorges, Blaum,
  Fr\"{o}mmgen, Geppert, Hammen, Kaufmann, Kr\"{a}mer, Krieger, Neugart,
  S\'{a}nchez et~al.}}]{GorgesJPB2015}
\bibinfo{author}{\bibfnamefont{C.}~\bibnamefont{Gorges}},
  \bibinfo{author}{\bibfnamefont{K.}~\bibnamefont{Blaum}},
  \bibinfo{author}{\bibfnamefont{N.}~\bibnamefont{Fr\"{o}mmgen}},
  \bibinfo{author}{\bibfnamefont{C.}~\bibnamefont{Geppert}},
  \bibinfo{author}{\bibfnamefont{M.}~\bibnamefont{Hammen}},
  \bibinfo{author}{\bibfnamefont{S.}~\bibnamefont{Kaufmann}},
  \bibinfo{author}{\bibfnamefont{J.}~\bibnamefont{Kr\"{a}mer}},
  \bibinfo{author}{\bibfnamefont{A.}~\bibnamefont{Krieger}},
  \bibinfo{author}{\bibfnamefont{R.}~\bibnamefont{Neugart}},
  \bibinfo{author}{\bibfnamefont{R.}~\bibnamefont{S\'{a}nchez}},
  \bibnamefont{et~al.}, \bibinfo{journal}{J. Phys. B: At. Mol. Opt. Phys.}
  \textbf{\bibinfo{volume}{48}}, \bibinfo{pages}{245008}
  (\bibinfo{year}{2015}).

\bibitem[{\citenamefont{Solaro et~al.}(2020)\citenamefont{Solaro, Meyer,
  Fisher, Berengut, Fuchs, and Drewsen}}]{SolaroPRL2020}
\bibinfo{author}{\bibfnamefont{C.}~\bibnamefont{Solaro}},
  \bibinfo{author}{\bibfnamefont{S.}~\bibnamefont{Meyer}},
  \bibinfo{author}{\bibfnamefont{K.}~\bibnamefont{Fisher}},
  \bibinfo{author}{\bibfnamefont{J.~C.} \bibnamefont{Berengut}},
  \bibinfo{author}{\bibfnamefont{E.}~\bibnamefont{Fuchs}}, \bibnamefont{and}
  \bibinfo{author}{\bibfnamefont{M.}~\bibnamefont{Drewsen}},
  \bibinfo{journal}{Physical Review Letters} \textbf{\bibinfo{volume}{125}}
  (\bibinfo{year}{2020}), ISSN \bibinfo{issn}{1079-7114},
  \urlprefix\url{http://dx.doi.org/10.1103/PhysRevLett.125.123003}.

\bibitem[{\citenamefont{Dubost et~al.}(2014)\citenamefont{Dubost, Dubessy,
  Szymanski, Guibal, Likforman, and Guidoni}}]{DubostPRA2014}
\bibinfo{author}{\bibfnamefont{B.}~\bibnamefont{Dubost}},
  \bibinfo{author}{\bibfnamefont{R.}~\bibnamefont{Dubessy}},
  \bibinfo{author}{\bibfnamefont{B.}~\bibnamefont{Szymanski}},
  \bibinfo{author}{\bibfnamefont{S.}~\bibnamefont{Guibal}},
  \bibinfo{author}{\bibfnamefont{J.-P.} \bibnamefont{Likforman}},
  \bibnamefont{and} \bibinfo{author}{\bibfnamefont{L.}~\bibnamefont{Guidoni}},
  \bibinfo{journal}{Phys. Rev. A} \textbf{\bibinfo{volume}{89}},
  \bibinfo{pages}{032504} (\bibinfo{year}{2014}),
  \urlprefix\url{https://link.aps.org/doi/10.1103/PhysRevA.89.032504}.

\bibitem[{\citenamefont{Dzuba et~al.}(1987)\citenamefont{Dzuba, Flambaum,
  Silvestrov, and Sushkov}}]{Dzuba1987_CorPotMethod}
\bibinfo{author}{\bibfnamefont{V.~A.} \bibnamefont{Dzuba}},
  \bibinfo{author}{\bibfnamefont{V.~V.} \bibnamefont{Flambaum}},
  \bibinfo{author}{\bibfnamefont{P.~G.} \bibnamefont{Silvestrov}},
  \bibnamefont{and} \bibinfo{author}{\bibfnamefont{O.~P.}
  \bibnamefont{Sushkov}}, \bibinfo{journal}{Journal of Physics B: Atomic and
  Molecular Physics} \textbf{\bibinfo{volume}{20}}, \bibinfo{pages}{1399}
  (\bibinfo{year}{1987}),
  \urlprefix\url{https://doi.org/10.1088/0022-3700/20/7/009}.

\bibitem[{\citenamefont{Agbemava et~al.}(2014)\citenamefont{Agbemava,
  Afanasjev, Ray, and Ring}}]{DDPC1}
\bibinfo{author}{\bibfnamefont{S.~E.} \bibnamefont{Agbemava}},
  \bibinfo{author}{\bibfnamefont{A.~V.} \bibnamefont{Afanasjev}},
  \bibinfo{author}{\bibfnamefont{D.}~\bibnamefont{Ray}}, \bibnamefont{and}
  \bibinfo{author}{\bibfnamefont{P.}~\bibnamefont{Ring}},
  \bibinfo{journal}{Phys. Rev. C} \textbf{\bibinfo{volume}{89}},
  \bibinfo{pages}{054320} (\bibinfo{year}{2014}),
  \urlprefix\url{https://link.aps.org/doi/10.1103/PhysRevC.89.054320}.

\bibitem[{\citenamefont{Raeder et~al.}(2018)\citenamefont{Raeder, Ackermann,
  Backe, and et~al}}]{No-IS-exp}
\bibinfo{author}{\bibfnamefont{S.}~\bibnamefont{Raeder}},
  \bibinfo{author}{\bibfnamefont{D.}~\bibnamefont{Ackermann}},
  \bibinfo{author}{\bibfnamefont{H.}~\bibnamefont{Backe}}, \bibnamefont{and}
  \bibinfo{author}{\bibnamefont{et~al}}, \bibinfo{journal}{Phys. Rev. Lett.}
  \textbf{\bibinfo{volume}{120}}, \bibinfo{pages}{232503}
  (\bibinfo{year}{2018}).

\end{thebibliography}

\end{document}